\documentclass[aps,pre,twocolumn,amsmath,amssymb,superscriptaddress]{revtex4-1}

\usepackage{graphicx}
\usepackage{dcolumn}
\usepackage{bm}
\usepackage{caption,subcaption}
\usepackage{hyperref}
\usepackage[usenames,dvipsnames]{color}

\newcommand{\fig}[1]{\textbf{Fig.~\ref{#1}}}
\newcommand{\eq}[1]{\textbf{Eq.~\ref{#1}}}
\newcommand{\sctn}[1]{\textbf{\S~\ref{#1}}}
\renewcommand{\vec}[1]{\underline{#1}}
\newcommand{\tens}[1]{\underline{\underline{#1}}}
\newcommand{\unity}{\tens{\hat{1}}}
\newcommand{\del}{\vec{\nabla}}

\newcommand{\kb}{k_\textmd{B}}
\newcommand{\kbt}{\kb T}

\newcommand{\av}[1]{\left\langle #1 \right\rangle}

\newcommand{\abs}[1]{\left| #1 \right|}
\newcommand{\transpose}[1]{{ #1 }^{T}}

\newcommand{\ie}{\textit{i.e.} }

\newcommand{\dt}{\delta t}

\newcommand{\vcm}[1]{\vec{v}_{\textmd{cm},#1}}
\newcommand{\rfric}{\gamma_R}
\newcommand{\suscHI}{\chi_\textmd{HI}}

\newcommand{\fmb}{f_\textmd{vel}}
\newcommand{\fms}{f_\textmd{ori}}
\newcommand{\wmf}{w_{\textmd{MF},c}}

\begin{document}
\title{Multi-Particle Collision Dynamics Algorithm for Nematic Fluids}

\author{Tyler N. Shendruk}
\email{ tyler.shendruk@physics.ox.ac.uk}
\homepage{ http://tnshendruk.com/}
\author{Julia M. Yeomans}
\affiliation{The Rudolf Peierls Centre for Theoretical Physics, Department of Physics, Theoretical Physics, University of Oxford, 1 Keble Road, Oxford, OX1 3NP, United Kingdom}


\begin{abstract}
Research on transport, self-assembly and defect dynamics within confined, flowing liquid crystals requires versatile and computationally efficient mesoscopic algorithms to account for fluctuating nematohydrodynamic interactions. 
We present a multi-particle collision dynamics (MPCD) based algorithm to simulate liquid-crystal hydrodynamic and director fields in two and three dimensions. 
The nematic-MPCD method is shown to successfully reproduce the features of a nematic liquid crystal, including a nematic-isotropic phase transition with hysteresis in 3D, defect dynamics, isotropic Frank elastic coefficients, tumbling and shear alignment regimes and boundary condition dependent order parameter fields. 
\end{abstract}

\maketitle


\section{Introduction}\label{sctn:intro}

As a state of soft condensed matter with intermediate symmetries between highly ordered crystals and disordered fluids, nematic liquid crystals are both phenomenologically fascinating and commercially valuable. 
No longer are liquid crystals of interest only to those producing liquid crystal display technology; now scientists interested in microfabricated systems\cite{ohzono12}, microelectromechanical devices\cite{beeckman11}, composite materials\cite{saez05}, biosciences\cite{woltman07} and active gels\cite{prost15} are exploiting the unique properties of liquid crystals in novel applications\cite{lagerwall12}. 
Interest in complex geometries (such as confining geometries nanoconfined geometries\cite{garlea15}, topological microfluidics\cite{sengupta13,sengupta14a} and colloidal intrusions\cite{dontabhaktuni14,jose14}) require versatile mesoscopic algorithms that can account for non-trivial boundary conditions. 
Likewise research into ``hypercomplex liquid crystals''\cite{dogic14} and self-assembly\cite{skarabot08,bisoyi11} would benefit from efficient methods to simulate nematohydrodynamic baths for macromolecular and colloidal solutes. 

Such elaborate systems present a considerable challenge for traditional particle-based numerical methods. 
Lattice Monte Carlo simulations have been very successful in simulating nematic liquid crystals\cite{lebwohl72} and continue to be widely employed due to their computational frugality\cite{ranjkesh14,chiccoli15}. 
However, out-of-equilibrium dynamics and relaxation mechanisms require more computationally costly methods. 
Off-lattice simulations of hard anisotropic particles and soft pair-potentials have played an important role in understanding generic liquid crystalline phases\cite{wilson09,frezza13}, but are limited to simple systems. 
Molecular dynamics simulations can account for molecular detail with a range of coarse-graining\cite{peter08,brini13}, including fully atomistic\cite{berardi04}, generic molecules\cite{hughes08} and the mesoscopic approach of dissipative particle dynamics\cite{levine05,lintuvuori08}. 
Even mesoscopic simulations can become computationally expensive when large numbers of constituent particles are required and so are generally limited to simplified systems. 

Investigating hypercomplex fluids or dynamics within demanding geometries calls for the continued development of versatile and computationally efficient coarse-grained algorithms. 
One mesoscopic simulation technique that has shown promising capabilities in simulating fluctuating hydrodynamics of isotropic solvents is the \emph{multi-particle collision dynamics} (MPCD) algorithm\cite{malevanets99,malevanets04}. 
MPCD has been used to simulate hydrodynamic interactions between macromolecules\cite{winkler04,jiang13} colloids\cite{radu14,poblete14}, vesicles\cite{noguchi05} and swimmers\cite{elgeti10,zottle14,schaar14}. 
It has even been extended to simulate viscoelastic fluids\cite{kowalik13} and electrohydrodynamics\cite{hickey12}. 
In this work, we propose an extension to the MPCD method to efficiently simulate fluctuating nematohydrodynamics (\emph{nematic-MPCD}).

\section{Method}\label{sctn:method}

Multi-particle collision dynamics algorithms forgo simulating molecular-scale interactions between constituent molecules. 
Instead, the continuum description is discretised into many artificial, point-like \emph{MPCD particles} that stochastically exchange momentum while respecting conservation laws for mass, momentum and energy. 
This is sufficient to reproduce the hydrodynamic equations of motion on sufficiently long length and time scales. 
Mesoscopic MPCD algorithms can dramatically reduce computational costs compared to simulations that explicitly calculate molecular pair-potentials and are well suited to simulating flowing systems involving non-trivial boundary conditions\cite{reid12,nikoubashman13}, finite Reynolds numbers\cite{prohm14}, and fluctuating hydrodynamics, which are ideal for moderate P\'{e}clet number systems\cite{padding04,shendruk13}. 

Here, we develop a nematic-MPCD method to efficiently simulate fluctuating nematohydrodynamics, by assigning an orientation pseudo-vector to each MPCD point-particle and updating orientations through a local and stochastic nematic \emph{multi-particle orientation dynamics} (MPOD) operator. 
Backflow and shear-alignment dynamics are ensured by coupling the MPCD and MPOD operators. 
In \sctn{sctn:results}, we demonstrate that nematic-MPCD reproduces the necessary physical properties to simulate a nematic liquid crystal when the velocity and director fields are coupled. 
In a very recent article, Lee and Mazza introduced an interesting hybrid, non-local MPCD method for liquid crystals~\cite{lee15}. 
The main difference to our approach is that their particles carry a director field that is coupled to the fluid through a discretisation of the stress terms in a simplified Ericksen-Leslie formalism of nematohydrodynamics. 

In this section, we begin by reviewing a traditional Andersen-thermostatted MPCD algorithm that conserves angular momentum. 
We go on to describe the implementation of the MPOD operator for nematic fluids and the two-way coupling between the director and velocity fields. 
Finally, we describe how potentially complex boundary conditions can be implemented. 

\subsection{Traditional MPCD for Isotropic Fluids}\label{sctn:mpcd}

The fundamental insight of MPCD algorithms is that continuous mass and momentum fields can be discretised into MPCD point-particles (labelled $i$). 
Each MPCD particle possesses a position $\vec{r}_i$, mass $m_i$ and velocity $\vec{v}_i$, and which interact through multi-particle, near-equilibrium stochastic collision events within lattice-based cells (labelled $c$) defined by a size $a$, population $N_c$, centre of mass velocity  centre $\vcm{c}=\av{ \vec{v}_j }_{N_c}$ and moment of inertia $\tens{I}_{c}=\sum_k^{N_c}m_k\left({r^\prime_k}^2\unity - \vec{r}^\prime_{k}\vec{r}^\prime_{k}\right)$ of the point-particles in cell $c$ relative to their centre of mass $\vec{r}_{\textmd{cm},c}$ where $\vec{r}_i^\prime=\vec{r}_i-\vec{r}_{\textmd{cm},c}$. 

The MPCD algorithms consist of two steps. 
Each MPCD particle streams ballistically for a time $\dt$ such that its position at time $t+\dt$ becomes 
\begin{align}
 \vec{r}_i\left(t+\dt\right) &= \vec{r}_i\left(t\right) + \vec{v}_i\left(t\right) \dt . 
 \label{eq:pos}
\end{align}
Multiple particles then undergo collision events, in which momentum is transferred between MPCD particles. 
To exchange momentum, the simulation domain is partitioned into cubic cells of thermally varying number density $\rho_c=N_c/a^d$ in $d$-dimensions. 
Discretising space into MPCD cells breaks Galilean invariance, though this can be remedied by randomly shifting the cell grid at each time step\cite{ihle01}. 
The collision operator $\Xi_{i,c}$ is a non-physical exchange designed to be stochastic and also to conserve the net momentum within each cell $c$, 
\begin{align}
 \vec{v}_i\left(t+\dt\right) &= \vcm{c}\left(t\right) + \Xi_{i,c}. 
 \label{eq:vel}
\end{align}

Many choices for the collision operator exist, which result in different versions of MPCD, including the original Stochastic Rotation Dynamics\cite{malevanets99,malevanets00} and a Langevin version of the algorithm\cite{noguchi07}. 
In this work, we utilise the Andersen-thermostatted collision operator\cite{noguchi07}
\begin{align}
 \Xi_{i,c} &= \vec{\xi}_i - \av{ \vec{\xi}_j }_{N_c} + \left( \tens{I}_c^{-1} \cdot \delta \vec{L}_c \right)\times\vec{r}_i^\prime , \label{eq:col}
\end{align}
where $\vec{\xi}_i$ is a random velocity drawn from the Maxwell-Boltzmann distribution $\fmb\left(\xi,\kbt\right)$ for thermal energy $\kbt$ and $\av{ \vec{\xi}_j }_{N_c}$ is the average of the $N_c$ random velocity vectors in the $c^\textmd{th}$ cell during the instant of the collision event. 
Randomly generating the $\vec{\xi}_i$ from the equilibrium distribution $\fmb$ in the moving reference frame ensures that the algorithm is locally thermostatted\cite{noguchi07}. 
The third term in the collision operator is a correction included to remove the angular momentum introduced by the collision operator 
\begin{align}
 \delta \vec{L}_c &= \sum_j^{N_c} m_j\left\{ \vec{r}^\prime_j \times \left(\vec{v}_j-\vec{\xi}_j\right)\right\} .
 \label{angMom}
\end{align}
Though the nematic-MPCD method does not strictly depend on this choice for $\Xi_{i,c}$, coupling the velocity field to the director field is accomplished by respecting this conservation law (see \sctn{sctn:angMomCons}). 

\subsection{Multi-Particle Orientation Dynamics for Nematic Fluids}\label{sctn:mpod}

We now that propose nematic liquid crystals can be simulated via a nematic-MPCD algorithm by including an orientation field. 

Each MPCD particle is assigned an orientation $\vec{u}_i$, while each cell acquires a tensor order parameter
\begin{align}
 \tens{Q}_{c} &= \frac{1}{d-1}\av{ d \vec{u}_i\vec{u}_i - \unity}_{N_c} .
 \label{eq:order}
\end{align}
For a nematic fluid, the largest eigenvalue is the local scalar order parameter $S_c$ of the cell and the local Frank director $\vec{n}_c$ is parallel to the corresponding eigenvector. 

Orientations interact through a positive, globally specified interaction constant $U$. 
In physical liquid crystals, the energy $U$ represents inter-molecular interactions and will be a non-constant function of temperature or molecular details such as nematogen dimensions and density. 
In this nematic-MPCD algorithm, the interaction constant $U$ is the simulation specified energy that governs the local evolution of orientations. 
Taking inspiration from the Andersen-thermostatted MPCD collision operator, we implement a stochastic multi-particle orientation dynamics operator for orientation. 
The essential requirements are that the MPOD operator must be \emph{local} and \emph{near equilibrium}, with no gradient terms in the collision operator. 
Therefore, we propose the orientation collision event
\begin{align}
 \vec{u}_i\left(t+\dt\right) &= \Psi_c\left( U,\tens{Q}_c\left(t\right) \right),
 \label{eq:ori}
\end{align}
where the multi-particle orientation operator $\Psi_c$ generates a random orientation $\vec{u}_i\left(t+\dt\right)$ drawn from the equilibrium probability distribution $\fms\left(U,\tens{Q}_c\left(t\right)\right)$ about the local director $\vec{n}_c\left(t\right)$ calculated from the tensor order parameter. 

\subsubsection{Maier-Saupe Distribution:\;}

As in the traditional Andersen-thermostatted MPCD algorithm, the multi-particle orientation operator depends on the condition of local, near-equilibrium statistics. 
In this work, we assume that the local equilibrium distribution for the orientation field obeys the Maier-Saupe self-consistent mean-field theory and so is an exponential function of $u_n \equiv \vec{u}_i\cdot\vec{n}_c$: 
\begin{align}
 \fms\left(U,S_c,\vec{n}_c\right) &= g e^{-\beta \wmf\left(U,S_c,u_n\right)} , 
 \label{eq:maiersaupe}
\end{align}
where $g$ is a normalisation constant, $\beta\equiv1/\kbt$ and each cell's \emph{mean-field interaction potential} is 
\begin{align}
 \wmf &= - U S_c u_n^2 + \frac{U}{d}\left(S_c-1\right). 
 \label{eq:wmf}
\end{align} 
The second term does not depend on $u_n$ and so the distribution of $u_n$ is determined by $e^{\beta US_c u_n^2}$. 
When the scaled energy $\beta US_c$ is small, all orientations are equally likely but when $\beta US_c$ is large the distribution becomes sharply oriented about $\vec{n}_c$. 

\subsubsection{Generating the Maier-Saupe Distribution:\;}
When $\beta US_c\approx1$, a Metropolis algorithm for $\wmf$ generates the random orientations. 
However, the distribution can be more efficiently approximated in the limits of $\beta US_c\gg1$ and $\beta US_c\ll1$. 
In the strong mean field limit $\beta US_c\gg1$, $f_\mathrm{ori}$ is sharply centred about $\vec{n}_c$ such that $u_n^2 = \cos^2\theta_n\approx1-\theta_n^2$, which means that the distribution for $\theta_n$ can be approximated as Gaussian $\fms \sim e^{ - \beta US_c\theta_n^2}$. 
The Gaussian approximation is used when $\beta US_c>5$. 
On the other hand, when $\beta US_c\ll1$ the exponent can be expanded and the cumulative distribution function of $\fms$ can be approximated as 
$W = \int_{-\infty}^{u_n} d\mu e^{\beta US_c\mu^2} \approx u_n + \beta US_c u_n^3 / 3$. 
Random values of $u_n$ can be generated through the transformation $u_n = 2^{-1/3} \kappa / \beta US_c - 2^{1/3} / \kappa$, where $\kappa\left(r\right) = \left( 3r\left(\beta US_c\right)^2+\left[9\left(\beta US_c\right)^4r^2+4\left(\beta US_c\right)^3\right]^{1/2} \right)^{1/3}$ and $r\in\left[0,1\right]$. 
This expansion is used when $\beta US_c<0.5$. 

\subsection{Two-way Coupling}\label{sctn:coupling}
Coupling between the director and fluid flow is crucial for reproducing nematohydrodynamics since flows can rotate the nematogens (\sctn{sctn:align}) and the rotation of nematogens in turn produces hydrodynamic motion, referred to as \emph{backflow} (\sctn{sctn:angMomCons}). 
We model the coupling to be superimposable overdamped torques. 

\subsubsection{Shear Alignment: Velocity$\rightarrow$Orientation Coupling:\;}\label{sctn:align}
The nematogens respond to the flow field's vorticity $\tens{\omega}=\left[\del\vec{v}-\transpose{\left(\del\vec{v}\right)}\right]/2$ and shear rate $\tens{D}=\left[\del\vec{v}+\transpose{\left(\del\vec{v}\right)}\right]/2$ (\fig{fig:period}; insets) by obeying the discretised Jeffery's equation for a slender rod
\begin{align}
 \frac{\delta\vec{u}_{\textmd{HI},i}}{\dt} &= \suscHI\left[ \vec{u}_i\cdot\tens{\omega}+ \lambda\left(\vec{u}_i\cdot\tens{D}-\vec{u}_i\vec{u}_i\vec{u}_i:\tens{D}\right) \right] ,
 \label{eq:jeff}
\end{align}
where $\lambda$ is the \emph{bare tumbling parameter} and $\suscHI$ is the \emph{shear coupling coefficient}, a simulation parameter that tunes the alignment relaxation time relative to $\dt$. 

For the rotation of an individual ellipsoidal particle subject to shear flow, $\suscHI=1$. 
When the shear coupling coefficient is set to zero ($\suscHI=0$) there is no coupling of the director to the velocity field. 

\subsubsection{Backflow: Orientation$\rightarrow$Velocity Coupling:\;}\label{sctn:angMomCons}
The nematogens are implicitly envisioned as rotating through the viscous fluid that they themselves represent, and hence they experience viscous torques. Assuming that the rotational motion is overdamped, we model this as a rotational Stokes drag.
In the nematic-MPCD algorithm, backflow coupling is accounted for by balancing this drag on the nematogens via transferring an equal and opposite change in angular momentum to the velocity collision operator. 

The magnitude of the rotational Stokes drag felt by the nematogens is $\vec{\Gamma}_i = \rfric \vec{u}_i\times \vec{\dot{u}}_i$. 
The nematic collision causes the MPCD point-particles to change their orientation by $\delta \vec{u}_{\textmd{col},i} = \vec{u}_i\left(t+\dt\right) - \vec{u}_i\left(t\right)$ over the time step $\dt$, and so results in a collisional contribution to the drag torque 
\begin{align}
 \vec{\Gamma}_{\textmd{col},i} &= \frac{\rfric}{\dt} \vec{u}_i\times \delta\vec{u}_{\textmd{col},i} .
 \label{eq:colTorque}
\end{align}
Likewise, the torque on the fluid due to the shear alignment is $\vec{\Gamma}_{\textmd{HI},i} = \rfric \vec{u}_i\times \delta\vec{u}_{\textmd{HI},i} /\dt$ from \eq{eq:jeff}. 
If any additional torques due to external fields were to be included they too would need to be included in the net torque. 

To balance these torques with the hydrodynamic drag, the opposite of the net change in the angular momentum $\delta\vec{\mathcal{L}}_c=\sum_i^{N_c} \delta\vec{\mathcal{L}}_i = \sum_i^{N_c}\Gamma_i \dt$ is transferred to the linear momentum portion of the algorithm. 
The MPCD collision operator $\Xi_{i,c}$ (\eq{eq:col}) is thus modified to account for liquid crystal backflow becoming 
\begin{align}
 \Xi_{i,c} &= \vec{\xi}_i - \av{ \vec{\xi}_j }_{N_c} + \left( \tens{I}_c^{-1} \cdot \left[ \delta\vec{L}_c-\delta\vec{\mathcal{L}}_c\right] \right)\times\vec{r}_i^\prime . 
 \label{eq:coupled}
\end{align}
In this way, the total angular momentum of the system is conserved and the orientation-velocity coupling is accounted for. 
By setting $\rfric=0$, the transferred angular momentum of each particle is zero $\delta\vec{\mathcal{L}}_i=0$ and this coupling can be turned off. 

\begin{figure}[tb]
\centering
  \includegraphics[width=0.5\textwidth]{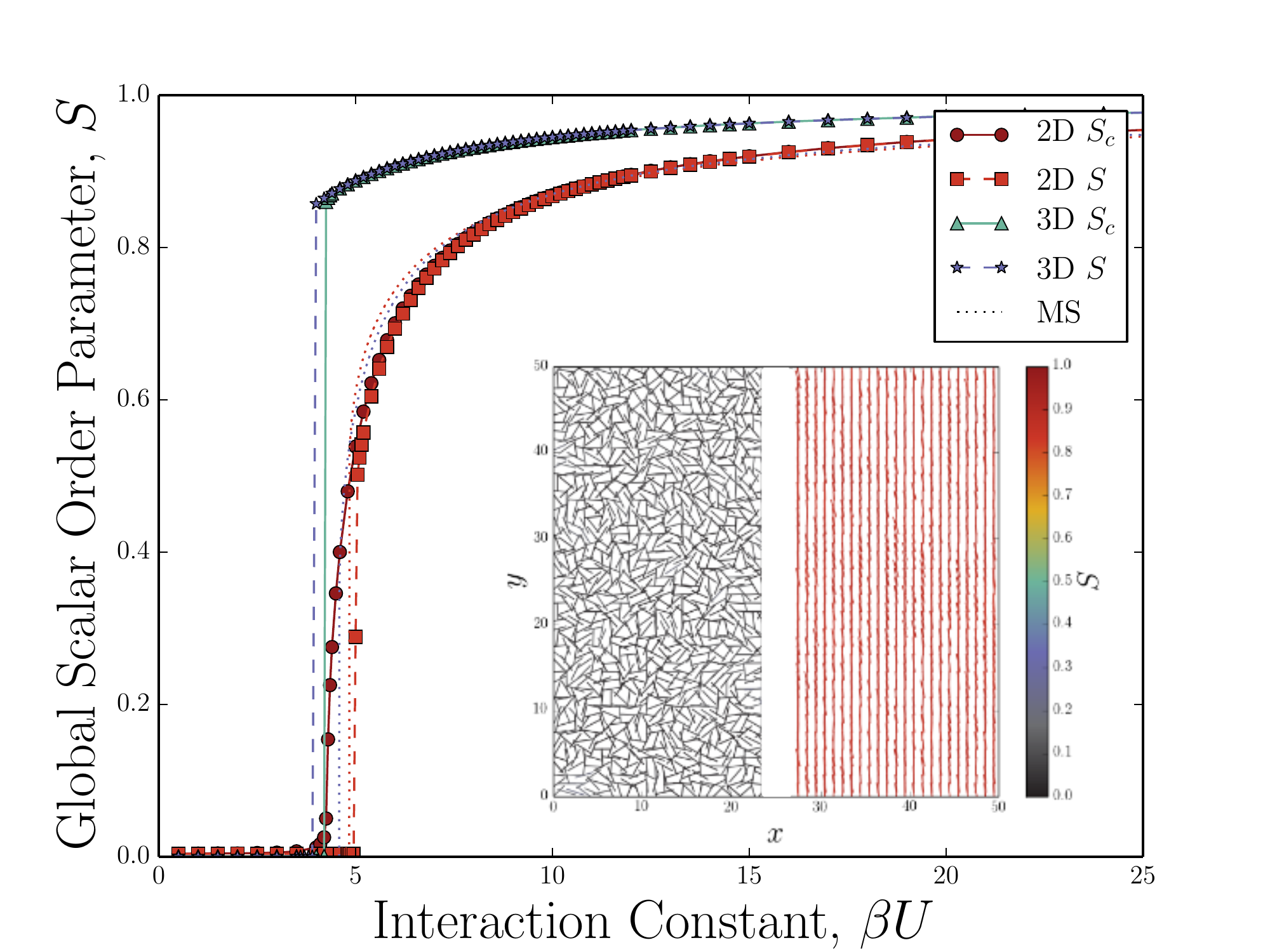}
  \caption{Nematic-isotropic phase transition. 
  Simulation parameters from \sctn{sctn:standard} are used with no shear coupling, $\suscHI=0$. 
  Simulations in 3D exhibit discontinuous isotropic to nematic transitions, regardless of whether the local $S_c$ (solid lines) or global $S$ (dashed lines) order parameter is used. 
  The transition is second order in 2D when the local $S_c$ is used but becomes discontinuous if $S$ is used. 
  The nematic-isotropic transition agrees qualitatively with the Maier-Saupe self-consistent mean-field theory (MS; dotted lines). 
  Inset shows a typical snapshot of the isotropic disordered state (left) and nematic ordered state (right). 
  }
  \label{fig:int}
\end{figure}

\subsection{Boundary Conditions}\label{sctn:bcs}
One of the advantages of particle-based hydrodynamics solvers is that complex and mobile boundary conditions can be implemented. 
For this reason, the nematic-MPCD may be well-suited to nematic fluids confined within microfluidic devices\cite{sengupta13,sengupta14a} and to simulating colloidal-liquid crystals\cite{dontabhaktuni14,jose14} and hypercomplex liquid crystals\cite{dogic14}. 

The effect of boundaries on positions and velocities are implemented in the standard manner. 
Periodic boundary conditions are implemented by wrapping the MPCD particle positions. 
Lees-Edwards boundary conditions are used to introduce simple shear flows across periodic domains\cite{kikuchi03}. 
No-slip walls are simulated by implementing bounce-back boundary conditions with phantom particles\cite{lamura02,whitmer10,bolintineanu12}. 

The boundary conditions also set the \emph{easy direction} describing the preferred orientation of the liquid crystal director at a surface. 
During a bounce-back collision event with the surface, the orientation $\vec{u}_i$ of the impinging nematic-MPCD particle is set parallel to the surface's easy direction. 
This anchoring is not strong, as will be seen in \sctn{sctn:wall}. 
For homeotropic boundary conditions, the easy direction is normal to the surface. 
For planar boundary conditions, the easy axis is parallel to the surface. 
In this case, all in-plane directions can be equivalent or a single preferred direction can be specified. 
If no preferred direction is specified then the boundary is said to be \emph{non-anchoring}. 

\subsection{Units and Chosen Simulation Parameters}\label{sctn:standard}
Values are expressed in MPCD simulation units --- time, mass, energy and length are given respectively by time step $\dt$, particle mass $m$, thermal energy $\kbt$ and cell size $a=\dt\sqrt{\kbt/m}$. 
The new MPOD parameters are also stated these units. 
The interaction constant $U$ has units $\kbt$, while the rotational friction coefficient $\rfric$ has units $\kbt\;\dt$. 
Both the bare tumbling parameter $\lambda$ and the shear coupling coefficient $\suscHI$ are dimensionless. 

Except when otherwise stated, the simulations presented in this manuscript vary input parameters about the following set of values: 
In this manuscript, simulations are carried out in 2D ($d=2$) for a system of size $V=50^d$ with periodic boundary conditions and a mean number density of $\rho=\av{N_c}=20$. 
The MPCD particles are randomly initiated with positions from a uniform distribution, velocities from the Maxwell-Boltzmann distribution and aligned nematic orientations. 
Parameter values are chosen to be $m=1$, $\kbt=1$, $\dt=1$, $a=1$, $U=15$, $\rfric=0.01$, $\lambda=2$ and $\suscHI=1$. 

\section{Results}\label{sctn:results}

Having described the implementation of the nematic-MPCD algorithm, we now characterise the resulting properties of the liquid crystal. 
We first consider how the isotropic-to-nematic phase transition depends on the simulation parameters, particularly the heuristic shear coupling coefficient and number density of MPCD particles. 
We measure the nematic-isotropic hysteresis and explore the dynamics of the defect annihilation rate as the system orders. 
Elastic free energy drives defect annihilation and we measure the isotropic Frank elastic coefficients to be a linear function of the interaction constant. 
The response of the isotropic phase to an ordering wall is characterised. 

\subsection{Nematic-isotropic transition}\label{sctn:int}

\begin{figure*}[tb]
  \centering
  \includegraphics[width=1.0\textwidth]{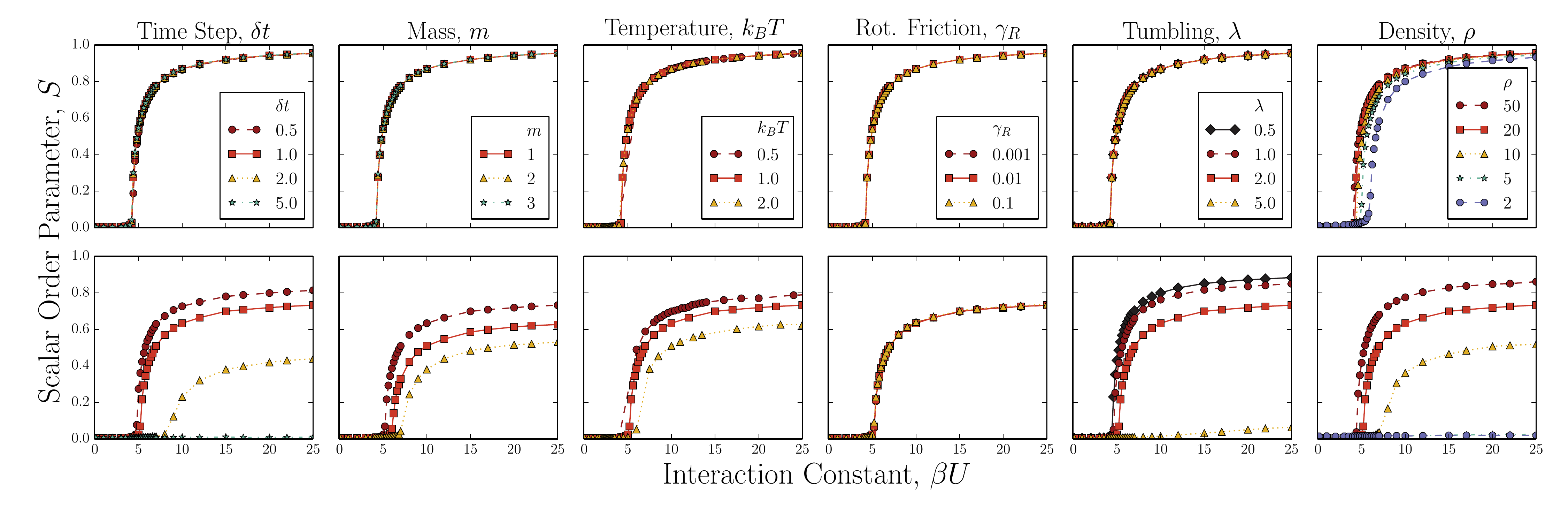}
  \caption{Global order parameter as a function of simulation parameters varied about the simulation values from \sctn{sctn:standard} (red squares). 
  The top row shows simulations in the absence of shear coupling, $\suscHI=0$. 
  The bottom row shows simulations with full coupling, $\suscHI=1$. 
  }
  \label{fig:varyParams}
\end{figure*}

When $\beta U$ is small, the nematic-MPCD algorithm exists as an isotropic fluid state with a small global order parameter $S$ (\fig{fig:int}; inset-left). 
When $\beta U$ is large a nematic state is formed (\fig{fig:int}; inset-right). 
Maier-Saupe self-consistent theory predicts that the nematic-isotropic transition is first order (\fig{fig:int}). 
Although the nematic-MPCD algorithm assumes near-equilibrium and so uses the Maier-Saupe distribution on the local cell level, the scalar order parameter and directors are spatially varying fields rather than mean-field values. 

In 3D systems of large enough size, periodic boundary conditions and no shear coupling, the nematic-MPCD algorithm does exhibit a strongly first order nematic-isotropic phase transition (\fig{fig:int}). 
In these simulations, the nematic fluid is initialised in the nematic state and resides in a periodic cube of size $50^3$. 
The system discontinuously jumps from zero to a global scalar order parameter $S^*=0.860\pm0.003$ at $\left[\beta U\right]^* = 4.20\pm0.05$. 

In 2D the nematic-isotropic transition is expected to become a Kosterlitz-Thouless-type transition\cite{stein78,vink14}. 
The present simulations demonstrate that the transition is no longer first order, increasing from zero at $\left[\beta U\right]^* = 4.1\pm0.1$ in a $50^2$ system (\fig{fig:int}). 
The second order nature of the nematic-isotropic transition is a direct result of the nematic-MPCD's ability to accommodate spatialtemporal varying fields. 
Future studies should more fully characterise the nature of the nematic-isotropic transition in 2D. 

\subsubsection{Global vs. Local Scalar Order Parameter:\;}
By replacing the local scalar order parameter $S_c$ of each cell $c$ with the system's globally determined order parameter $S$ in each cell's local mean-field interaction potential $\wmf$ (\eq{eq:wmf}), the 2D transition becomes first order (\fig{fig:int}). 
The order parameter curve remains relatively unchanged except near the phase transition. 
The transition from the isotropically disordered state is retarded compared to the spatially varying case that uses the local order parameters but suddenly  jumps to $S^*=0.51\pm0.01$ at $\left[\beta U\right]^* = 5.00\pm0.05$. 

\subsubsection{Variation of Simulation Parameters:\;}

In order to assess the impact of varying simulation parameters on the nematic-isotropic transition, we initially omit the velocity$\rightarrow$orientation coupling by setting $\suscHI=0$ in \eq{eq:jeff} (\fig{fig:varyParams}; top row). 
We consider varying time step $\dt$, mass $m$, temperature $\kbt$, rotational friction coefficient $\rfric$, bare tumbling parameter $\lambda$ and mean number density $\rho$. 
In the zero-coupling limit, \fig{fig:varyParams} (top row) shows that none of the MPCD simulation parameters have a significant effect on the nematic ordering. 
Only the mean number density has an observable affect on the curve. 
At extremely low mean number densities, the transition occurs at a slightly larger interaction constant. 
It should be noted that when an individual nematic-MPCD particle is alone in an MPCD cell neither its velocity nor its orientation are altered. 

\subsubsection{Impact of Coupling Fluctuating Hydrodynamics:\;}

When the shear coupling coefficient $\suscHI$ is zero the global scalar order parameter $S$ rises from zero in the isotropic phase to $S=1$ in the $\beta U\rightarrow\infty$ ordered limit. 
This is no longer true when $\suscHI\neq0$ (\fig{fig:varyParams}; bottom row). 
As the hydrodynamic coupling is restored by increasing $\suscHI$, the value of the scalar order parameter decreases for a given interaction constant. 
This occurs because fluctuations in the velocity field introduce an additional source of noise through \eq{eq:jeff} when $\suscHI\neq0$. 
These fluctuations reduce the order in the director field and move the system away from the fully ordered state of $S=1$. 
With full coupling, only the rotational friction coefficient $\rfric$ is seen to have no impact on the $S$ curve (\fig{fig:varyParams}; bottom row). 
This is because $\rfric$ controls the rotational relaxation dynamics and does not influence the equilibrium state. 

\begin{figure}[tb]
  \centering
  \includegraphics[width=0.5\textwidth]{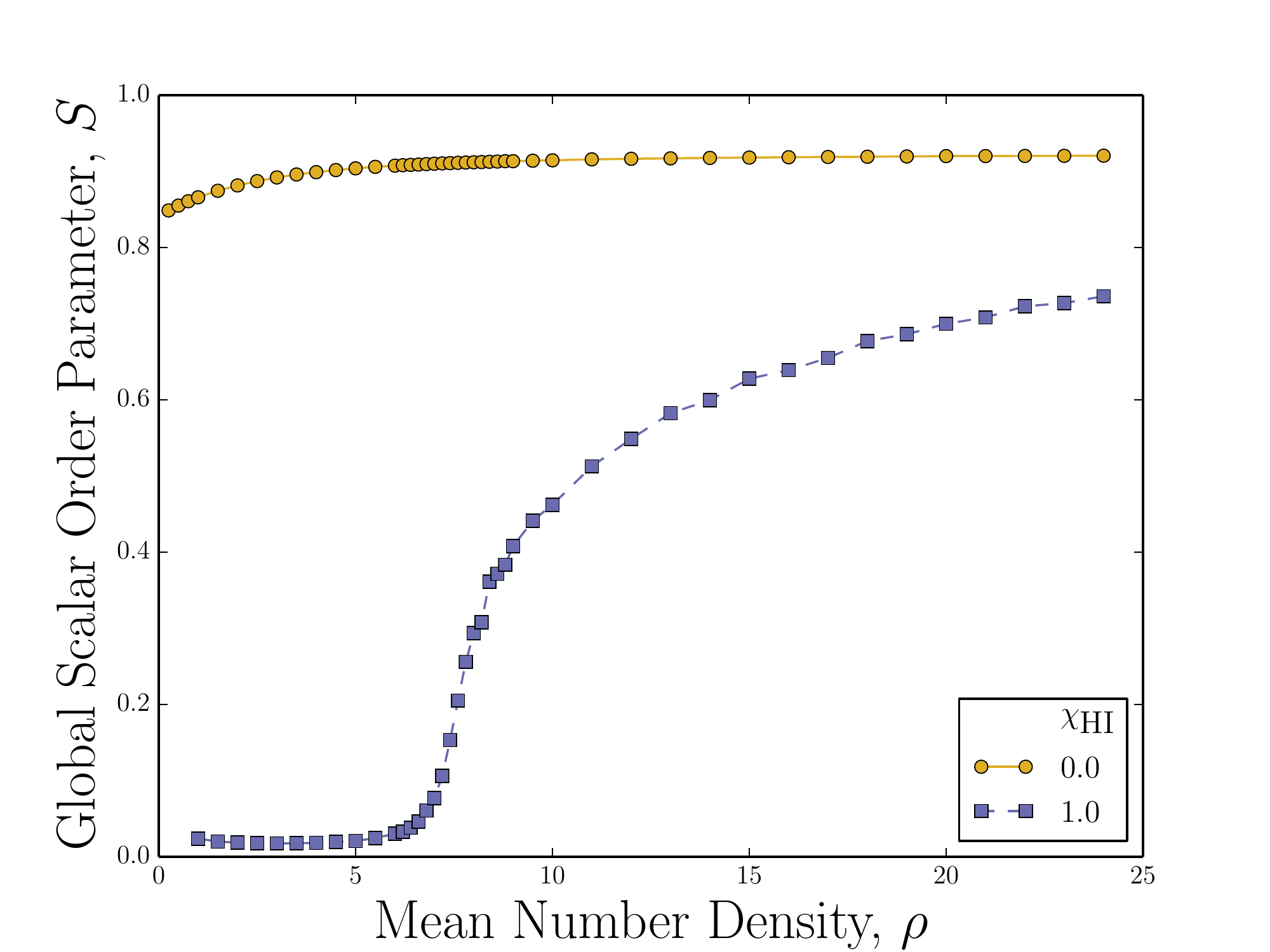}
  \caption{Nematic-isotropic phase transition as a function of average number density for $U=15$. 
  Simulation parameters from \sctn{sctn:standard} are used. 
  Without shear coupling ($\suscHI=0$), the global order parameter $S$ is relatively constant but when $\suscHI=1$ the system exhibits a density dependent second order phase transition. 
  }
  \label{fig:densTrans}
\end{figure}

When $\suscHI=0$, the mean number density is the only simulation parameter seen to have any observable effect on the isotropic to nematic transition and then only at extremely low values (\fig{fig:varyParams}; top row). 
At the lowest number density the transition is less sharp and occurs at a slightly higher interaction constant $\beta U$. 
While $S$ depends weakly on $\rho$ when $\suscHI=0$ (\fig{fig:densTrans}), it is a strong function of number density when $\suscHI=1$. 
\fig{fig:densTrans} shows the global scalar order parameter as a function of density for $U=15$. 
When the shear coupling parameter $\suscHI$ is set to zero, the system remains in the nematic state even at quite low densities. 
On the other hand, when coupling is included, $S$ increases from zero with mean number density. 
In fact, the order parameter in \fig{fig:densTrans} exhibits a continuous transition and the nematic-MPCD algorithm possesses a nematic-isotropic transition as a function of density when $\suscHI=1$. 

\begin{figure}[tb]
  \centering
  \includegraphics[width=0.5\textwidth]{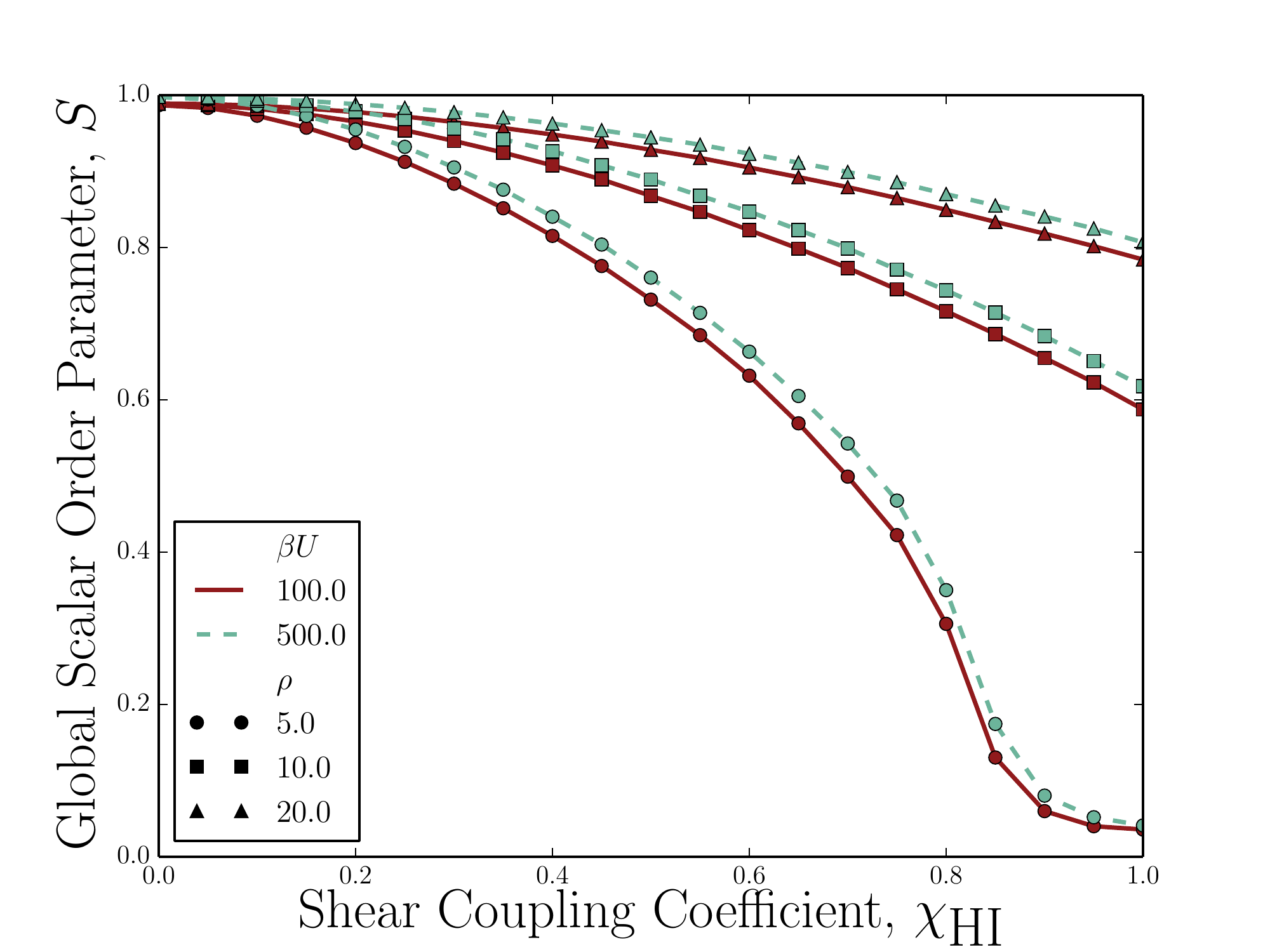}
  \caption{Increased shear coupling ($\suscHI$) reduces the scalar order parameter. 
  The order parameter is measured in the highly nematic phase at $\beta U=100$ and $500$. 
  The other simulation parameters are given in \sctn{sctn:standard}. 
  }
  \label{fig:coupling}
\end{figure}

Since there is a nematic-isotropic transition as a function of density (\fig{fig:densTrans}), it is clear that the shear coupling coefficient has a larger effect at lower number densities than it does at larger densities. 
\fig{fig:coupling} shows the strong interaction limit of $S$ (measured at $\beta U=100$ and $500$) for various densities as a function of coupling. 
For a low mean number density of $\rho=5$, \fig{fig:coupling} shows that the strong limit drops from $\lim_{\beta U\rightarrow\infty}S\approx1$ when $\suscHI=0$ to only $0.038\pm0.003$ when the algorithm is fully coupled. 
Fluctuations are pronounced because of the small number fluctuations of particles in each MPCD cell. 
By increasing the mean number density $\rho$, the continuum limit is approached and fluctuations become less severe. 
When $\rho=20$ and the algorithm is fully coupled ($\suscHI=1$), the strong interaction limit is $S=0.80\pm0.01$ (\fig{fig:coupling}). 
Throughout this work, we set the mean number density $\rho=20$, though a lower density may suffice in many situations. 

When the algorithm is fully coupled with $\suscHI=1$, the tumbling parameter can also increase the susceptibility of the order field to velocity fluctuations through \eq{eq:jeff}. 
This can be seen in \fig{fig:varyParams} (bottom row). 
When $\lambda=5$, fluctuations in the shear rate $\tens{D}$ fully disorder the system. 

\subsubsection{Hysteresis:\;}\label{sctn:hyst}

\begin{figure}[tb]
\centering
  \includegraphics[width=0.5\textwidth]{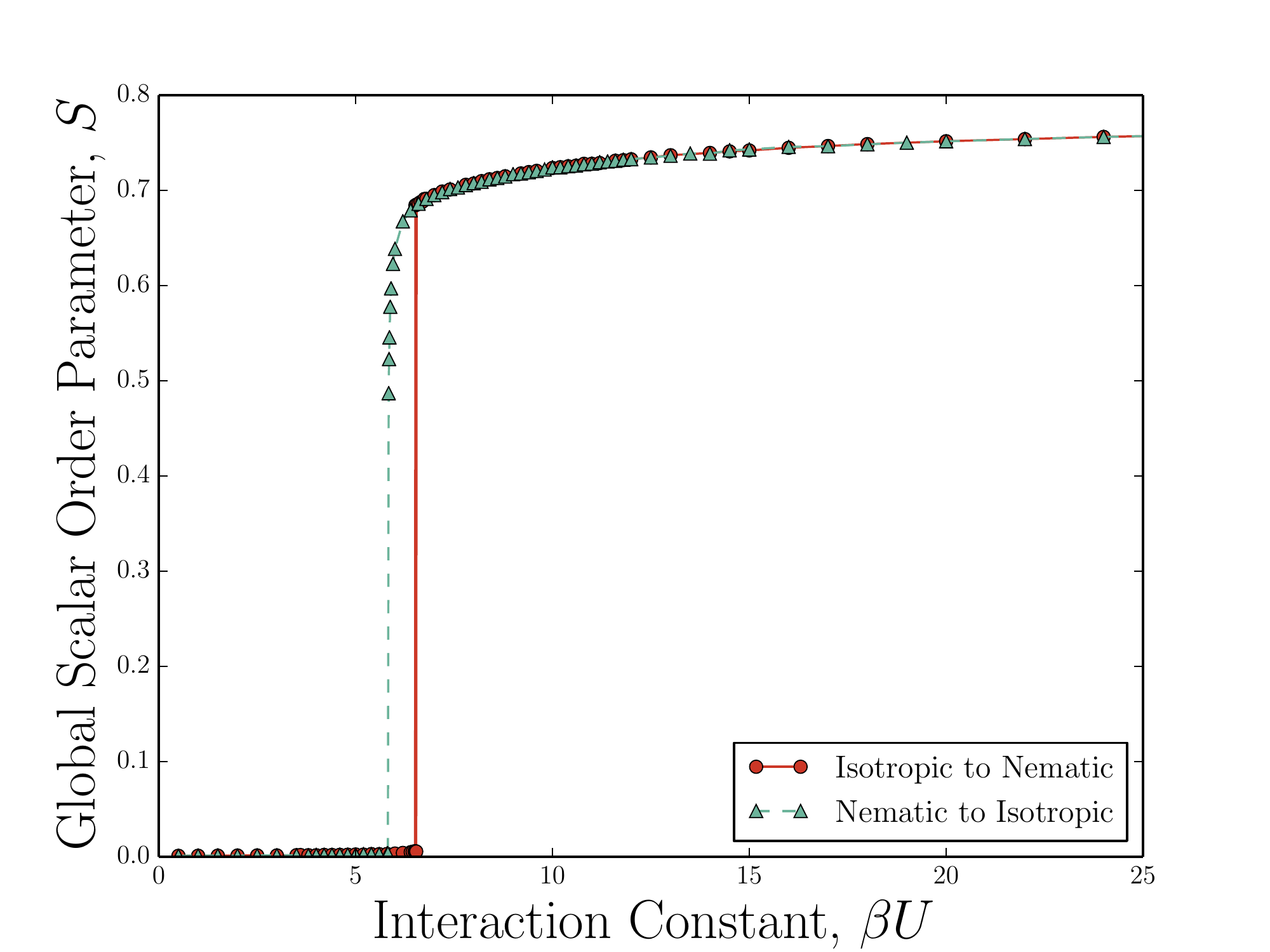}
  \caption{Hysteresis loop in the isotropic/nematic transition. 
  Simulations are initialized in the isotropic (circles) or nematic state (triangles). 
  Simulation parameters from \sctn{sctn:standard} are used in 3D. 
  }
  \label{fig:hyst}
\end{figure}

Hysteresis is expected in 3D due to the first order nature of the nematic-isotropic transition. 
By comparing 3D nematic-MPCD simulations initialised with the director field in the isotropic state (as in in \sctn{sctn:int}) to those initialised in the nematic state, a striking hysteresis loop is observed in \fig{fig:hyst}. 
The interaction constant, $\beta U$, is fixed throughout the duration of individual simulations. 
At these system sizes, the width of the hysteresis is measured from \fig{fig:hyst} to be $\beta\Delta U^* = 0.70\pm0.03$ and the difference in order parameters at the transition points is $\Delta S^*=0.20\pm0.04$. 

\subsection{Defect Annihilation Dynamics}

\begin{figure}[tb]
\centering
  \begin{subfigure}[t]{0.23\textwidth}
    \includegraphics[width=1.0\textwidth]{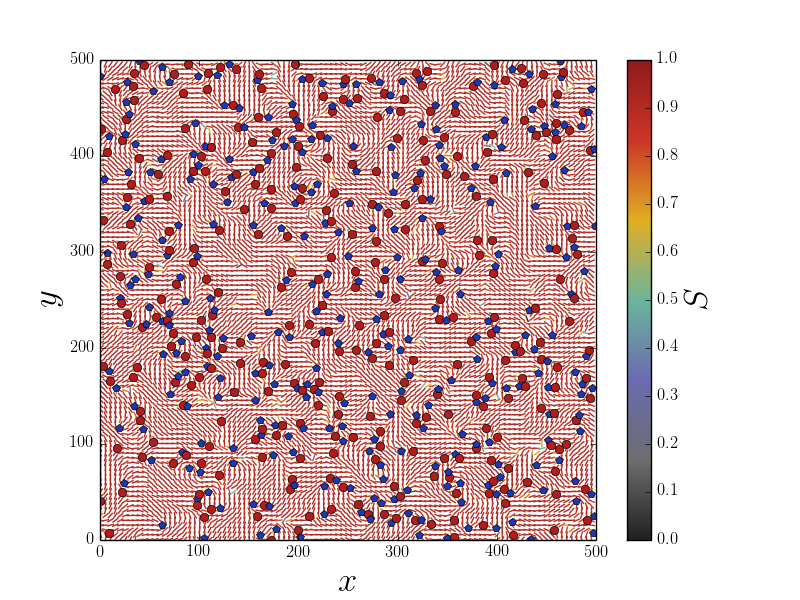}
    \caption{Director field at $t=40\dt$.}
    \label{fig:t40}
  \end{subfigure}
  \begin{subfigure}[t]{0.23\textwidth}
    \includegraphics[width=1.0\textwidth]{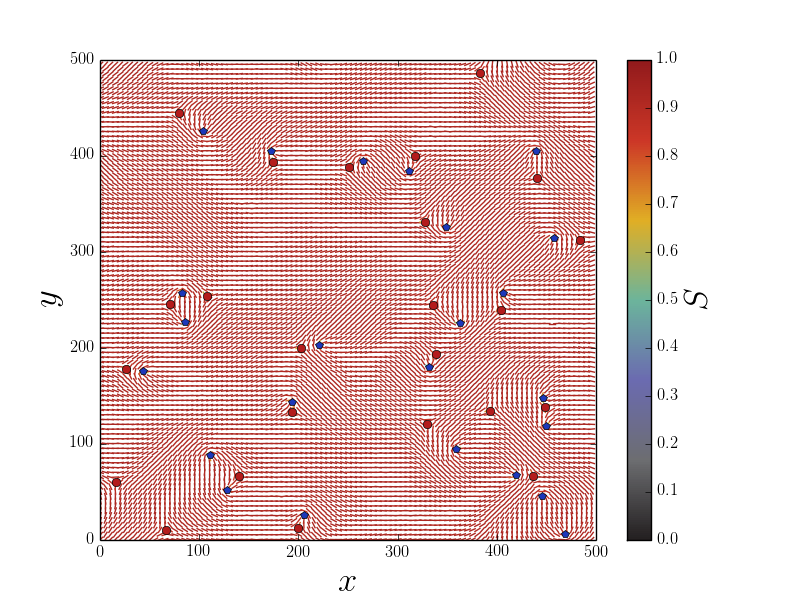}
    \caption{Director field at $t=400\dt$.}
    \label{fig:t400}
  \end{subfigure}
  \\
  \begin{subfigure}[t]{0.5\textwidth}
    \includegraphics[width=1.0\textwidth]{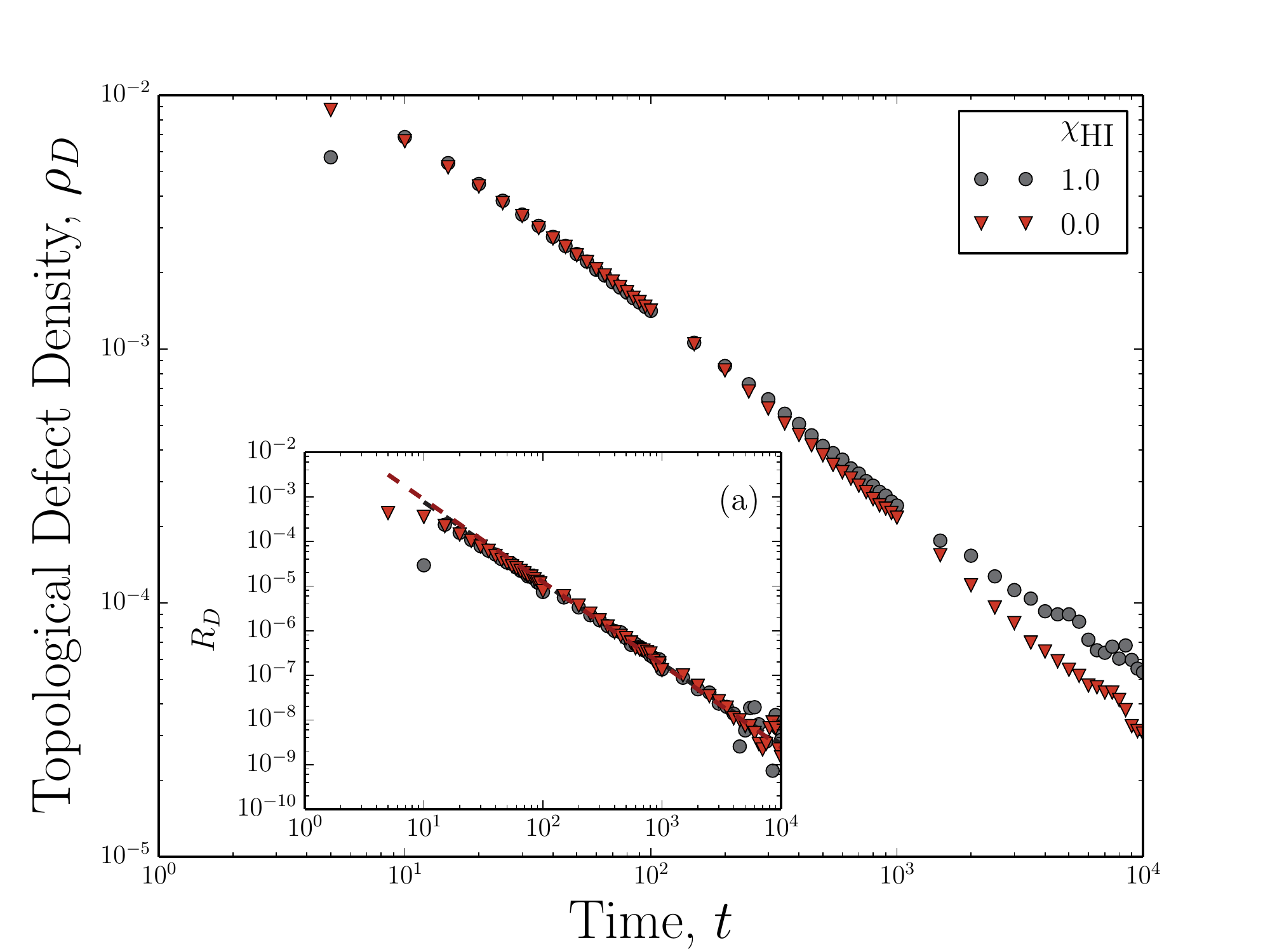}
  \end{subfigure}
  \caption{Number density of topological defects as a function of time with and without shear coupling. 
  \textbf{Inset (a):} Annihilation rate of defects $R_\textmd{D}$. 
  Simulation parameters from \sctn{sctn:standard} are used with a 2D system size of $500\times500$. 
  In the defect maps, colour denotes local scalar order parameter and defects are mapped with red circles marking $-1/2$ defects and blue pentagons marking $+1/2$ defects. 
  }
  \label{fig:defectPop}
\end{figure}

The process of transitioning from the isotropic to nematic phase discussed in \sctn{sctn:int} is controlled by the dynamics of topological defects. 
Though the increasing interaction constant $U$ generates local order along a spontaneous direction $\vec{n}_c$, neighbouring regions may break symmetry along any other direction. 
Therefore, many $\pm1/2$ topological defects rapidly emerge from the disordered director field. 
Pairs of oppositely charged defects must approach each other and annihilate for global ordering. 

Since the 2D number density $\rho_\textmd{D}=0.0080\pm0.0005$ of defects is initially quite high, the annihilation rate $R_\textmd{D}=-\dot{\rho}_\textmd{D}$ is large but falls rapidly (\fig{fig:defectPop}; inset a). 
As the density decreases, the average separation between topological defects increases and annihilation events become less frequent (compare \fig{fig:t40} showing an example system at $t=40$ to \fig{fig:t400} showing the same system at $t=400$). 
A variety of scaling relations for the annihilation rate $R_\textmd{D}\left(t\right) \sim t^{-\left(\nu+1\right)}$ have been put forward. 
Mean-field arguments predict $\nu=1$, purely diffusive kinetics suggest $\nu=0.5$ and scaling arguments give $\nu=6/7$\cite{liu97}. 
Furthermore, the scaling law possesses short-time logarithmic corrections\cite{zapotocky95,denniston01b}. 
When measured on short times between $t\in\left[10,10^3\right]$, the nematic-MPCD annihilation rate appears to decay as  $\nu=0.74\pm0.02$ (\fig{fig:defectPop}) but the exponent increases to $\nu=0.83\pm0.04$ when evaluated over $t\in\left[10^2,10^4\right]$ (\fig{fig:defectPop}), which is in agreement with the $\nu=6/7$ scaling prediction. 

\subsection{Frank Elastic Coefficients}\label{sctn:elastic}

\begin{figure}
  \centering
  \includegraphics[width=0.5\textwidth]{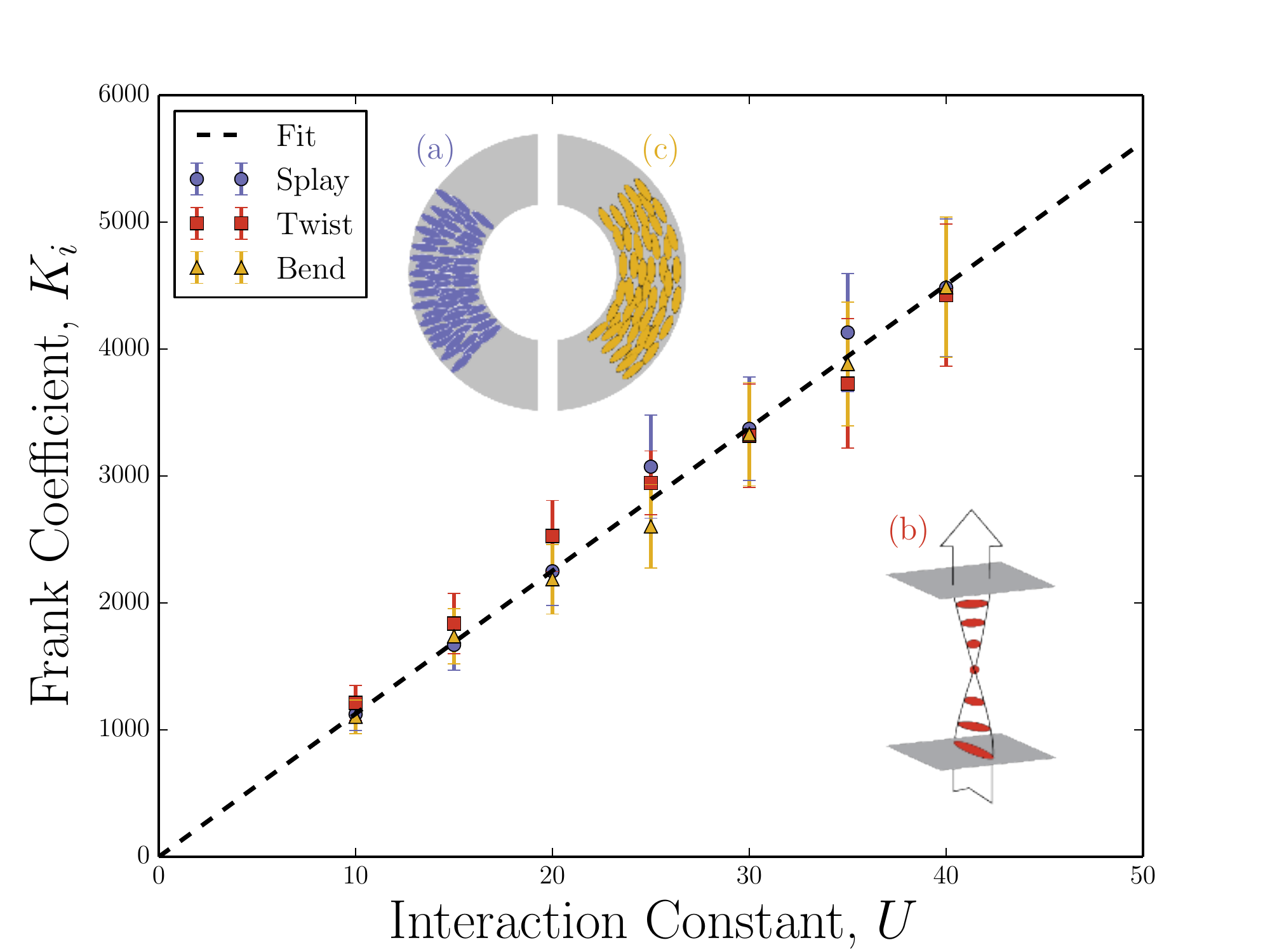}
  \caption{Frank elastic coefficients for splay, twist and bend as a function of interaction constant $U$. 
  Simulation parameters from \sctn{sctn:standard} are used for a system of size $30\times30\times30$ ($N=5.4\times10^5$ nematogens) with $\suscHI=0$.
  Insets a,b,c respectively depict splay, twist and bend.}
  \label{fig:elastic}
\end{figure}

We have considered how the nematic state arises from the isotropic state. 
Let us now consider the nematic response to distortions in the director field. 
Gradients in the director field $\vec{n}$ lead to the free energy density per unit volume $f = K_\textmd{splay} \left(\del\cdot\vec{n}\right)^2 / 2 + K_\textmd{twist} \left(\vec{n}\cdot\del\times\vec{n}\right)^2 / 2 + K_\textmd{bend} \left(\vec{n}\times \left(\del\times\vec{n}\right)\right)^2 / 2 $. 
Splay, bend and twist deformations are illustrate in \fig{fig:elastic}; insets. 
Since distortion are typically large compared to molecular length scales, the Frank elastic coefficients $K_\textmd{splay}$, $K_\textmd{twist}$ and $K_\textmd{bend}$ are macroscopic material properties. 


One technique for obtaining the Frank coefficients from particle-based simulations is to measure the equilibrium, orientational fluctuation spectrum~\cite{cleaver91,allen96,wilson05,gemunden15}. 
In reciprocal space, the tensor order parameter for each wave vector is $\tens{\hat{Q}}\left(\vec{k}\right) = \rho^{-1} \sum_{i=1}^{N} \tfrac{1}{d-1}\left( d \vec{u}_i\vec{u}_i - \unity \right)  \exp\left(i\vec{k}\cdot\vec{r}_i\right)$. 
We work in a varying director-based coordinate system, in which $\vec{n}_c=\left[0,0,1\right]$ and the wave vector is in the 13-plane, \ie $\vec{k}=\left[k_1,0,k_3\right]$. 
In this coordinate system, the equipartition theorem~\cite{cleaver91} relates the the low $\left|\vec{k}\right|$ limit of the orientational fluctuations to the Frank coefficients 
\begin{align}
 \av{ \hat{Q}_{\alpha 3}\left(\vec{k}\right) \hat{Q}_{\alpha 3}\left(-\vec{k}\right) } =  \frac{9}{4} \frac{S V \kbt}{K_\alpha k_1^2 + K_\textmd{bend} k_3^2}
 \label{eq:reciprocal}
\end{align}
for $\alpha=1,2$ (splay, twist). 
Through \eq{eq:reciprocal}, the Frank coefficients may be determined as fitting parameters of the fluctuation spectrum in reciprocal space. 
A large system size of $V=30\times30\times30$ and $N=5.4\times10^5$ MPCD particles are used in the following simulations to ensure sufficient statistics for many near-zero $\left|\vec{k}\right|$ values and accurate fits. 

The resulting Frank coefficients of the nematic-MPCD fluid are shown in \fig{fig:elastic}. 
Although splay, twist and bend deformations may possess differing coefficients in some physical systems, simple scaling suggests that all three elastic constants are of order $\sim U/a$ and theoretical considerations of the Maier-Saupe self-consistent model\cite{marrucci91} predict $K_i=\ell^2\rho US^2\left[1+C_i\right]/6$, where $i=\left\{\textmd{splay}, \textmd{twist}, \textmd{bend}\right\}$ and $\ell$ is a characteristic interaction distance. 
The different constants $C_i$ depend on the molecular details and higher moments of the orientation distribution\cite{marrucci91}. 
The nematic-MPCD simulations ostensibly exhibit isotropic elasticity. 
This is expected because, in the limit that the rod length is small compared to the interaction length, the constants $C_i$ are safely neglected and the Frank coefficients are predicted to converge\cite{marrucci91}. 
Since the nematic-MPCD algorithm simulates point-like nematogens with a characteristic interaction length equal to the finite cell size, the \emph{one-constant approximation} applies. 

In agreement with simple scaling and the Maier-Saupe self-consistent  predictions, the measured elastic coefficients for the nematic-MPCD algorithm are linear with respect to the interaction constant $U$ (\fig{fig:elastic}). 
Together, they are fit to $K_i = \left(113\pm5\right)U$ for the parameters in \fig{fig:elastic}. 

\subsection{Tumbling and Shear Alignment}\label{sctn:flowAlign}
Thus far, we have considered quiescent nematic fluids. 
We now turn our attention to flowing systems. 
Microscopically, the director field is influenced by shearing flows through \eq{eq:jeff} and as described schematically in \fig{fig:period}; inset. 

In the infinitely dilute limit of a suspension of spheroidal particles, the bare tumbling parameter is a geometrical entity that can be cleanly related to the particle aspect ratio $p$ by $\lambda=\left(p^2-1\right)/\left(p^2+1\right)$, which goes to unity as $p\rightarrow\infty$ and is zero for spheres ($p=1$). 
However, interactions between nematogens in a nematic fluid allow the actual tumbling parameter to deviate from the isolated-slender-rod value and distributions of molecules can exhibit effective tumbling parameters that are larger than unity. 
Such fluids are referred to as aligning-nematics because there is a stable alignment angle, the Leslie angle $\theta_L$, between the director and shear field. 

By considering a Fokker-Planck equation for the probability distribution of orientations, Archer and Larson\cite{archer95} found that the flow tumbling behaviour of ellipsoidal particles with $\lambda=\left(p^2-1\right)/\left(p^2+1\right)$ is determined by the tumbling parameter
\begin{align}
 \lambda^\prime &= \lambda\frac{15S+48S_4+42}{105S}. 
 \label{eq:tumble}
\end{align}
In the nematic-MPCD algorithm, $\lambda$ is the specified simulation parameter for the bare tumbling parameter, the magnitude of which can be set larger than unity. 
We shall see that $\lambda^\prime$ as given by \eq{eq:tumble} is the resulting tumbling parameter of the nematic-MPCD algorithm. 
In \eq{eq:tumble}, $S_4$ is the fourth moment of the Maier-Saupe probability distribution. 
The distribution can be written as an expansion of orthogonal Gegenbauer polynomials $C^{(\gamma)}_n\left(x\right)$ in $d$-dimensions as
\begin{align}
 \fms\left(U,S,\vec{n}\right)  &= \sum_{\ell=0}^\infty \frac{4\ell+1}{2} \mathcal{S}_{2\ell} C^{(\frac{d-2}{2})}_{2\ell}\left(u_n\right),
\end{align}
where the moments are $\mathcal{S}_{2\ell} = \av{C^{(\frac{d-2}{2})}_{2\ell}}$. 
In 3D, the polynomials are Legendre polynomials, while they are Chebyshev polynomials in 2D. 
The first even moment is the scalar order parameter $S \equiv \mathcal{S}_{2} = \tfrac{d}{d-1} \left\langle u_n^2-\tfrac{1}{d} \right\rangle$ representing the variance of the alignments about the director, while $S_4 \equiv \mathcal{S}_{4}$ is the next non-zero moment. 

\subsubsection{Tumbling Nematic:\;}\label{sctn:tumble}
\begin{figure}[tb]
\centering
  \includegraphics[width=0.5\textwidth]{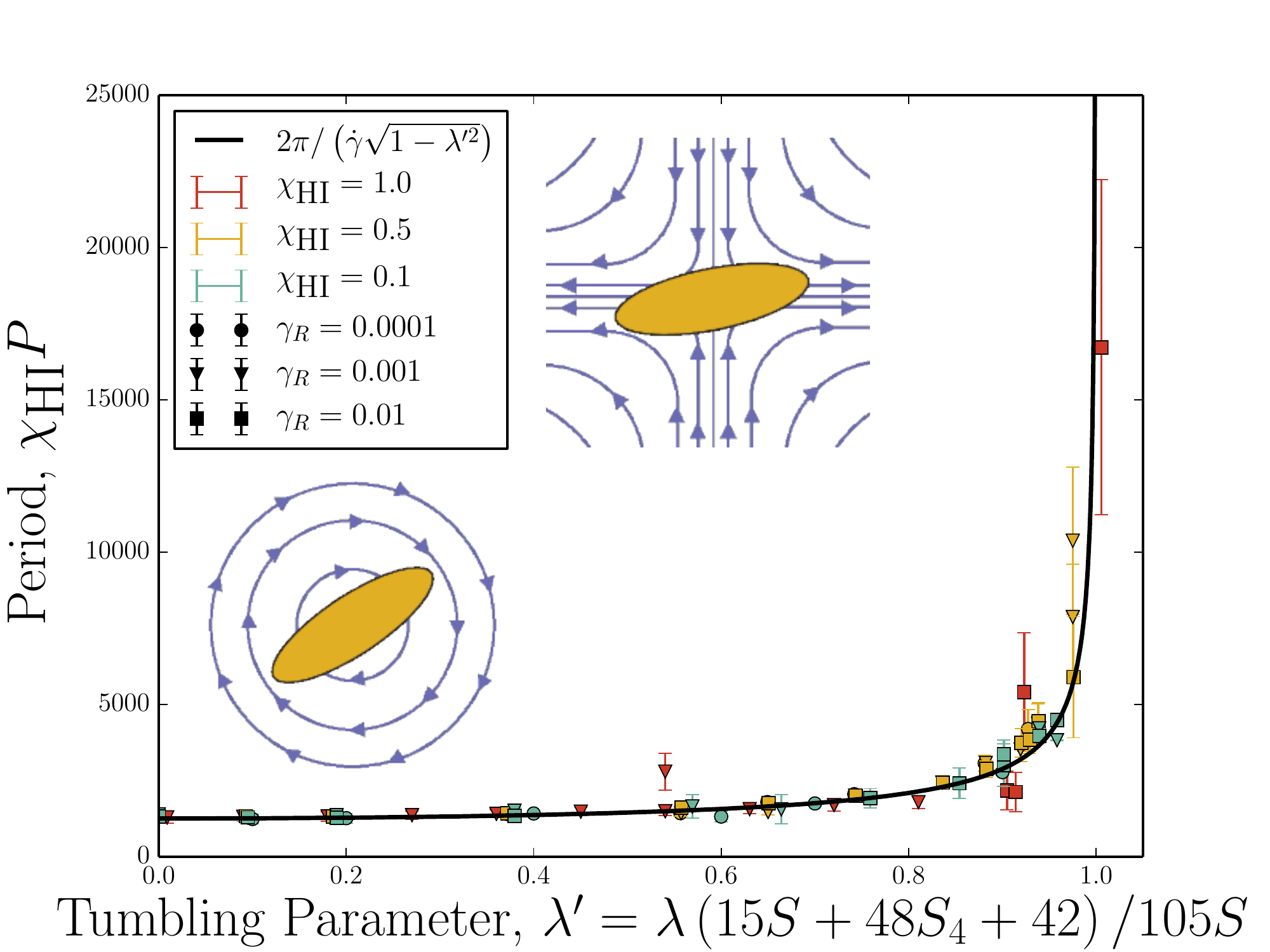}
  \caption{Jeffery periods corrected for the tumbling parameter and non-unity shear coupling coefficient. 
  Simulation parameters from \sctn{sctn:standard} are used with $U=20$ and orientations initialized in the nematic state. 
  The insets show the rotational and extensional components of \eq{eq:jeff}. 
  }
  \label{fig:period}
\end{figure}

When $\lambda^\prime<1$, the nematogens continuously revolve or \emph{tumble}. 
The tumbling period is set by the Jeffery orbits to be
\begin{align}
 P &= \frac{2\pi}{\suscHI \dot{\gamma}\sqrt{1-\lambda^{\prime2}}}, 
 \label{eq:period}
\end{align}
where $\dot{\gamma}$ is the shear rate. 

Using Lees-Edwards boundary conditions~\cite{kikuchi03} to establish a shear rate $\dot{\gamma}=0.01$ across a periodic channel of height $L=50$, we measure the tumbling period as a function of tumbling parameter $\lambda^\prime$ (\fig{fig:period}). 
The period is relatively small when $\lambda^\prime$ is small and varies very little as a function of tumbling parameter. 
However, as the tumbling parameter increases, the period increases rapidly and diverges as $\lambda^\prime\rightarrow1$. 
The simulated tumbling periods are found to be in good agreement with \eq{eq:period}. 

The tumbling period does not depend on the rotational friction coefficient $\rfric$ (\fig{fig:period}). 
This is expected both from inspection of \eq{eq:jeff} and from the realisation that the differential drag by the shearing flow is what rotates the rod. 

\subsubsection{Shear-Aligning Nematic:\;}

\begin{figure}[tb]
  \centering
  \includegraphics[width=0.5\textwidth]{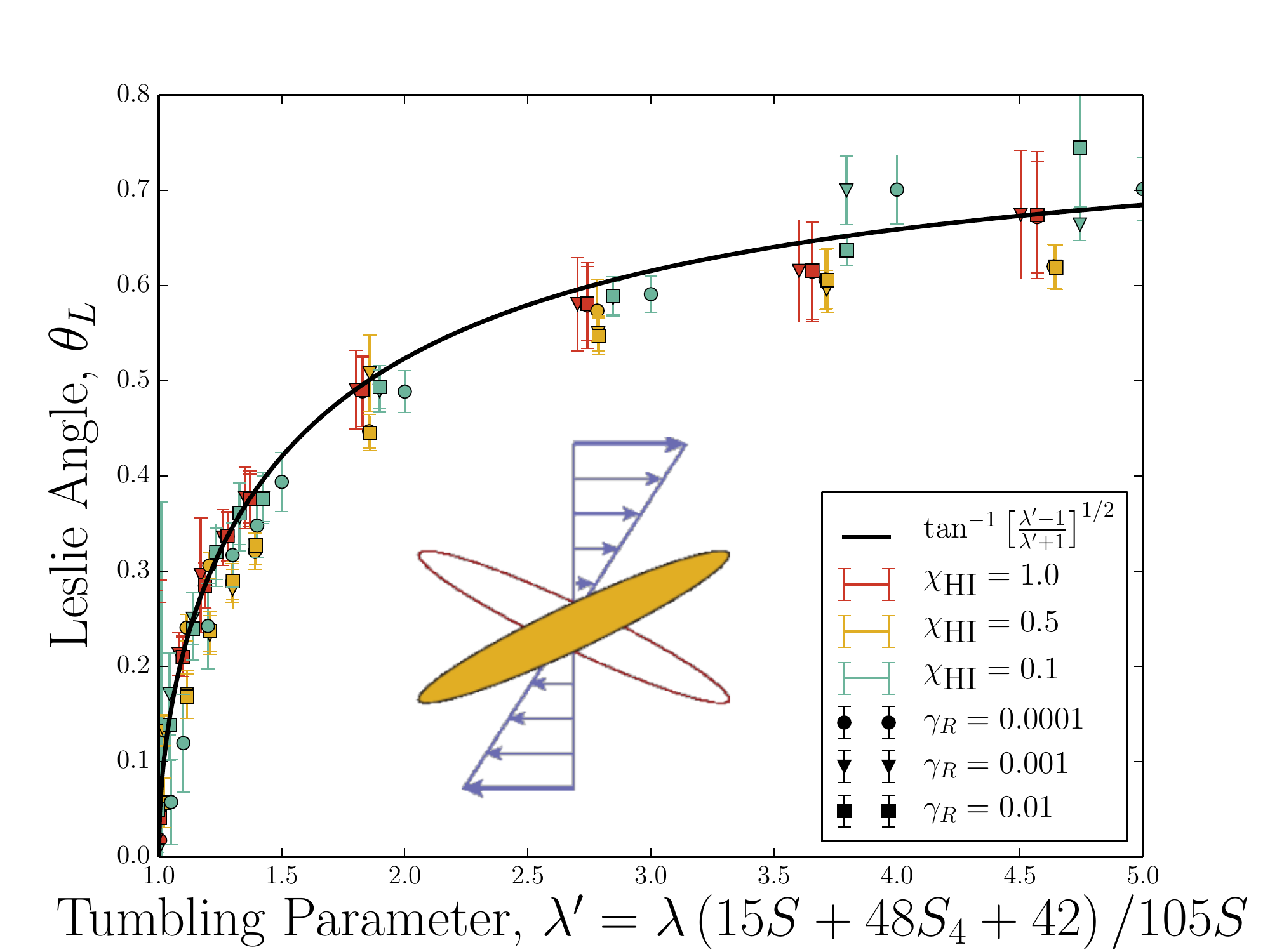}
  \caption{Leslie angles $\theta_L$ corrected for the tumbling parameter. 
  Simulation parameters from \sctn{sctn:standard} are used with $U=20$ and orientations initialized in the nematic state. 
  Ericksen diagram is inset and shows stable and unstable orientations for $\theta_L$. }
  \label{fig:leslie}
\end{figure}

When the magnitude of the bare tumbling parameter $\lambda$ is set so that $\abs{\lambda^\prime}$ is larger than unity, the nematogens do not tumble but rather align with the shear. 
For these tumbling parameters, \eq{eq:jeff} has the solution
\begin{align}
 \tan \theta_L &= \pm \sqrt{\frac{\abs{\lambda^\prime}-1}{\abs{\lambda^\prime}+1}}. 
 \label{eq:leslie}
\end{align}
Good agreement is found between \eq{eq:leslie} and the simulations using Lees-Edwards boundary conditions when the tumbling parameter (\eq{eq:tumble}) is greater than unity. 
As the tumbling parameter tends to $1^{+}$, the Leslie angle approaches zero. 
In this limit, the nematogens orient along the flow direction. 
When $\lambda^\prime \gg 1$, the Leslie angle of the nematic-MPCD fluid approaches $\pi/4$ as predicted by \eq{eq:leslie}. 

\subsection{Wall-Induced Ordering}\label{sctn:wall}

\begin{figure}
  \centering
  \includegraphics[width=0.5\textwidth]{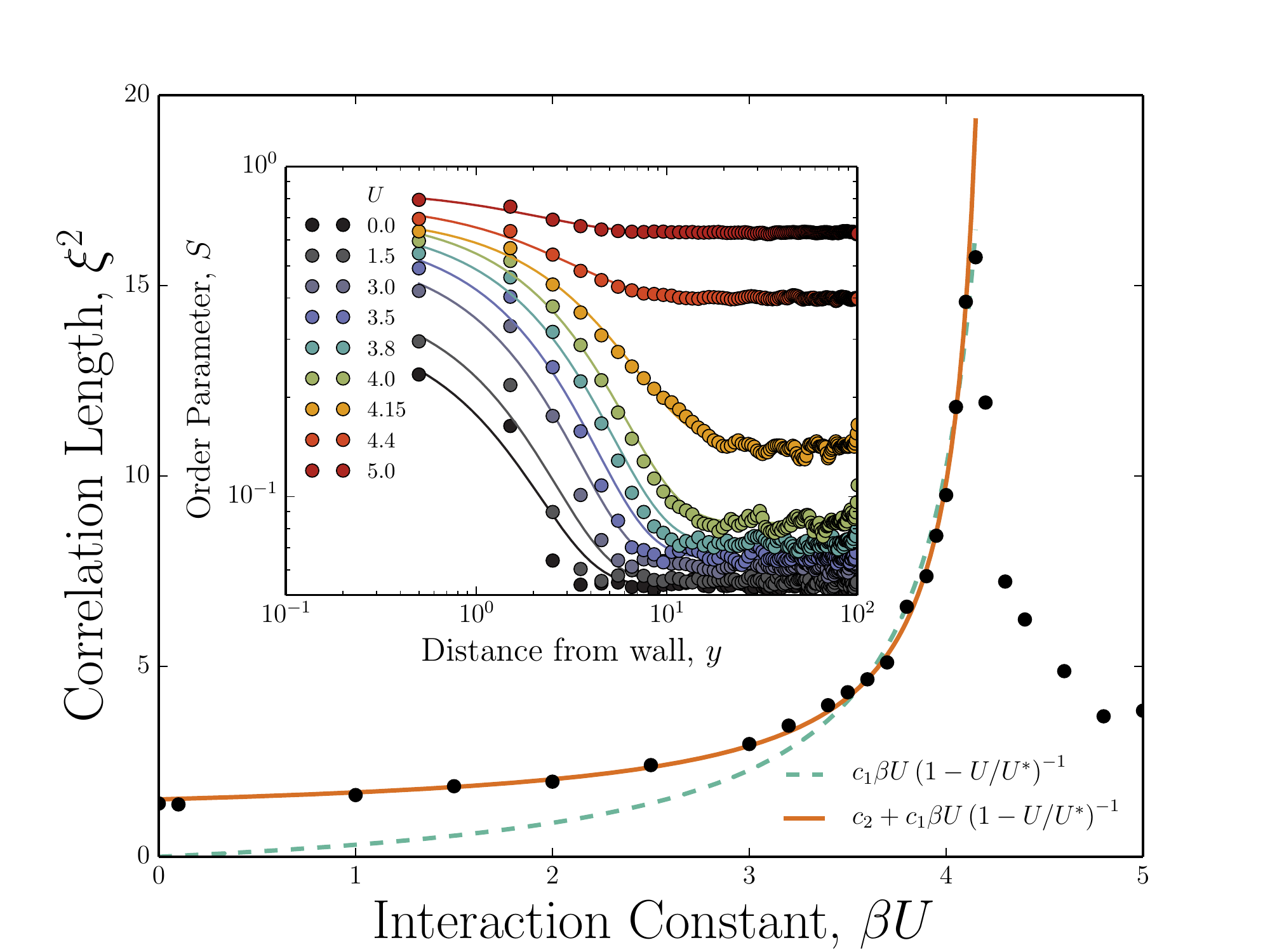}
  \caption{Wall-induced ordering. 
  Inset shows exponential decay of order from homeotropic wall towards the isotropic bulk, as characterised by the coherence length $\xi$. 
  Coherence length diverges as the nematic-isotropic transition is approached. 
  Simulation parameters from \sctn{sctn:standard} are used with $\rfric=1$. 
  }
  \label{fig:wallOrdering}
\end{figure}

Confining walls affect the nematic ordering. 
In the isotropic state, anchoring can cause ordering in the vicinity of the walls. 
We consider a 2D nematic-MPCD fluid confined between two no-slip plates separated by $L=100$. 
The plate at $y=0$ enacts homeotropic boundary conditions, which order the nematic fluid. 
The plate at $y=L$ is a non-anchoring boundary, which does not set a condition for $\vec{u}_i$. 

When the interaction constant is much less than the nematic-isotropic transition value $\left[\beta U\right]^*$ (\sctn{sctn:int}), the order decreases to the isotropic state far from the wall. 
As the interaction constant is increased, the value of the scalar order parameter $S_0\left(U\right)\equiv S\left(U,y=0\right)$ at the wall increases (\fig{fig:wallOrdering}; inset). 
This signifies that the anchoring is not infinitely strong and is strongly effected by the value of $U$. 

Additionally, the order extends further into the bulk fluid as $U$ increases. 
The characteristic distance the order extends from the wall is a coherence length $\xi$ (\fig{fig:wallOrdering}). 
One can predict that the order decays as $S\left(U,y\right)=S_0\left(U\right)e^{-y/\xi}$ by considering the total free energy functional to be the highest order term in the Landua-De~Gennes free energy and the deformation free energy. 
The coherence length is a function of the elastic constant and the distance from the transition, $\xi \propto \left( \tfrac{3K_\textmd{splay}+2K_\textmd{twist}}{T-T^*} \right)^{1/2}$. 
As the nematic-isotropic transition is approached, the coherence length diverges. 
The nematic-MPCD simulations accurately reproduce the exponential decay far below the transition point (\fig{fig:wallOrdering}; inset). 

It was seen in \sctn{sctn:elastic} that $K_i\sim U$ so the coherence length takes the form
\begin{align}
 \xi &= \left( \frac{c_1\beta U}{1-U/U^*} \right)^{1/2}. 
 \label{eq:wallcoherence}
\end{align}
The theory captures the rapid growth of the coherence length near the nematic-isotropic transition but goes to zero as $U\rightarrow0$, while the simulations do not (\fig{fig:wallOrdering}). 
The coherence length of the nematic-MPCD does not go to zero because MPCD algorithms are not able to resolve material properties on length scales comparable to the cell size $a$. 
If a second fitting parameter $c_2=\left(1.505\pm0.005\right)a$ is included as in \fig{fig:wallOrdering} to account for this discretisation effect, then \eq{eq:wallcoherence} well-represents the divergence of the coherence length in the isotropic phase. 

In the nematic phase, the order still decreases exponentially from $S_0$ but decays to a non-zero bulk value (\fig{fig:wallOrdering}; inset). 
Except near the nematic-isotropic transition, the order parameter falls steeply to its bulk value over a length scale comparable to a single MPCD cell. 

\section{Conclusions}\label{sctn:conc}

We have proposed a nematic-MPCD algorithm for simulating fluctuating nematohydrodynamics. 
Nematic-MPCD uses traditional Andersen-thermostatted MPCD with conservation of angular momentum to integrate the velocity field and a novel multi-particle orientation dynamics (MPOD) collision operator to progress the director field. 
By stochastically drawing orientations from the local Maier-Saupe equilibrium distribution, the MPOD operator updates the orientations without numerically evaluating gradients. 
In addition, the two-way coupling between the MPCD and MPOD operators represents backflow and shear-alignment. 
We have shown that this nematic-MPCD algorithm reproduces the essential physical properties of a simple nematic fluid, such as the nematic-isotropic phase transition, topological defects, Frank elasticity and shear alignment. 

The nematic-MPCD algorithm holds much promise as a tool for simulating nematohydrodynamics, but future studies should carefully investigate the anchoring strength (since modifications to the no-slip conditions were required in traditional MPCD\cite{lamura02,whitmer10,bolintineanu12}) and work towards kinetic theories to quantitatively predict the nematic material properties as a function of simulation parameters. 
Though simple, the algorithm holds exciting potential for simulating a wide variety of soft matter systems. For example defect dynamics within topological microfluidic devices\cite{sengupta14a} or porous media\cite{araki13} could be modelled, exploiting the ease with which the algorithm can handle complicated confining geometries.  It would also be of interest to consider dispersed nanoparticles, carbon fibres\cite{lagerwall08}  or swimmers\cite{zhou14}  within a liquid crystal host, and it is relatively easy to imagine that generalized Maier-Saupe theories\cite{greco14} could be implemented to in the MPOD collision operator to simulate cholesteric or biaxial liquid crystals. 
 
\section*{Acknowledgements}
This work was supported through EMBO funding to T.N.S (ALTF181-2013) and ERC funding to J.M.Y. (291234 MiCE).
We thank Jens Elgeti for early discussions and Amin Doostmohammadi for critical readings of the code and manuscript.

\bibliography{rsc}

\end{document}